\documentclass[dvips]{acta}
\usepackage{supertabular,lscape,epsfig}
\usepackage{amssymb}
\usepackage{amsmath}
\SetPages{00}{000}
\SetVol{59}{2009}

\begin{document}
\renewcommand{\thefootnote}{\fnsymbol{footnote}}
\setcounter{footnote}{1}
\def\deg{$^{\rm o}$}
\begin{Titlepage}
\Title{The All Sky Automated Survey.\\ The Catalog of Variable Stars in the Kepler Field of View}
\Author{A.~~P~i~g~u~l~s~k~i}
{Astronomical Institute, University of Wroc{\l}aw, ul.~Kopernika 11, 51-622 Wroc{\l}aw, Poland\\ 
e-mail:pigulski@astro.uni.wroc.pl}
\Author{G.~~P~o~j~m~a~\'n~s~k~i,~~~B.~~P~i~l~e~c~k~i~~~and~~~D.~M.~~S~z~c~z~y~g~i~e~{\l}}
{Warsaw University Observatory, 
Al.~Ujazdowskie~4, 00-478~Warszawa, Poland\\
e-mail: (gp,pilecki,dszczyg)@astrouw.edu.pl}
\Received{February 8, 2009}
\end{Titlepage}

\vspace*{-12pt}
\Abstract{We present the catalog of 947 variable stars located in the field of view of the Kepler satellite. The catalog is a result
of the analysis of $VI$ photometry obtained during the first 17-month observations in the ASAS3-North station. The variable stars we present
are divided into eleven groups according to the presented variability; the groups are briefly discussed. The catalog is 
intended to be a source of information for target selection process and follow-up programs.}
{Stars: binaries -- Stars: pulsations -- Stars: classification}

\Section{Introduction}
The main scientific goal of the NASA Kepler\footnote{The overview and status of the Kepler mission can be found at 
{\it http://kepler.nasa.gov/}} satellite mission (Borucki \etal 1997) is to search for extrasolar planets, 
especially in habitable zones, and to characterize the properties of the detected planetary systems. 
This will be achieved by means of a very precise photometry and transit detection. Over 10$^5$ targets
from a sample of stars of different mass and age will be selected for observation. The properties of both planets and their parent stars, 
especially stellar radii, can be determined in some cases when the stars exhibit $p$-mode pulsations. This is why
asteroseismic program has been included as a part of the Kepler mission (Christensen-Dalsgaard \etal 2007). 
In comparison with ground-based data, the advantage of Kepler observations will be a very high precision of photometry combined
with a very long uninterrupted sequences of data. Thus, a problem with aliasing, inherently associated with the
ground-based data, will be avoided. Due to the brightness range (9--15 mag in $V$) and location in the sky, Kepler observations of
pulsating stars will complement those of other satellite missions like MOST (Walker \etal 2003) and CoRoT 
(Barge \etal 2008).

\begin{figure}[!h]
\includegraphics[width=12.5cm]{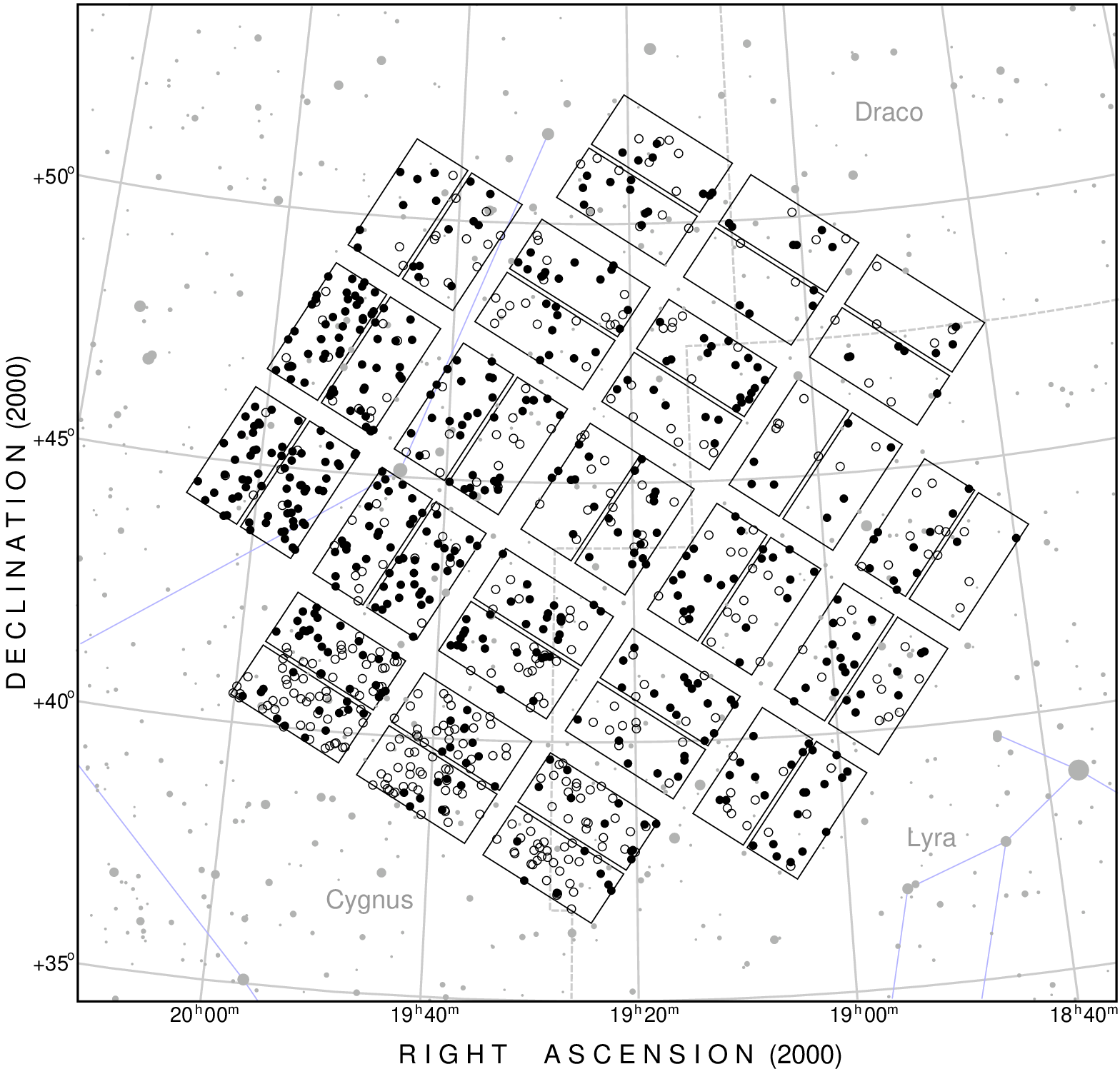}
\FigCap{The 18{\deg} $\times$ 18{\deg} map of the Kepler FoV located in Cygnus, Lyra and Draco. The rectangles delineate the limits of 42 Kepler CCDs.
The stars from our catalog are shown as filled (new) and open (already known variable stars) circles.}
\label{map}
\end{figure}

The Kepler forty two 2200\,$\times$\,1024-pixels CCD detectors will cover over a hundred square degrees in Cygnus, Lyra and Draco (see Fig.~\ref{map}). 
Every three months the telescope will be rotated around the optical axis by 90{\deg}. Because of the symmetry of the detectors, 
the rotation will not affect the coverage of the sky (with only a small exception of a tiny fraction of the observed
field close to the center of the Kepler field of view). In the course of the work on catalog which is presented in this paper, we considered
only those stars that fall within the Kepler field of view (FoV). Other variables, located between or outside Kepler CCD chips, are not considered here.
The catalog contains data for 947 variable and candidate variable stars and was released primarily because of the need of target selection for
the first part of the mission.

\begin{figure}[htb]
\includegraphics[width=12.5cm]{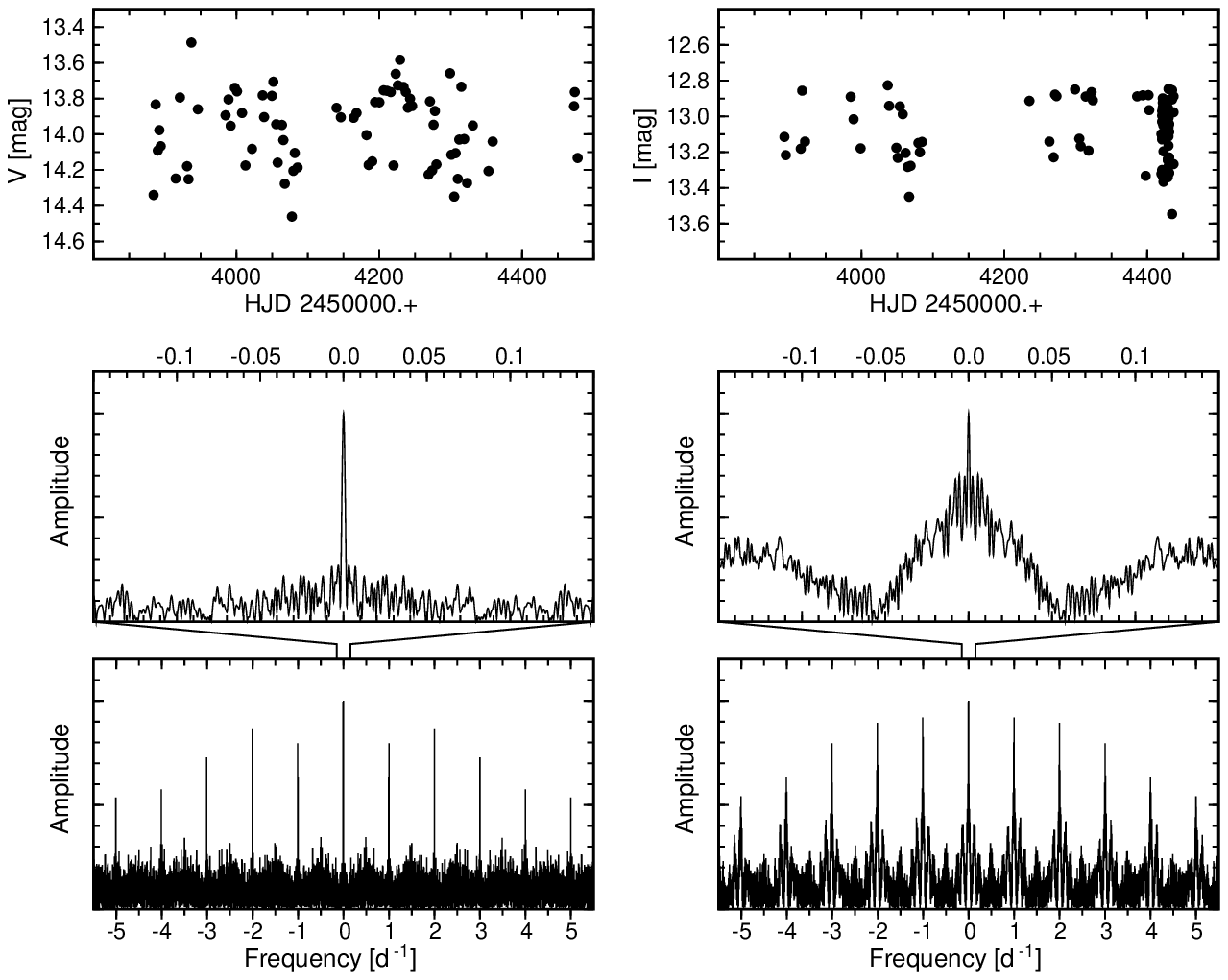}
\FigCap{{\it Upper panels:} Typical distribution of $V$ ({\it left}) and $I$ ({\it right}) ASAS data analyzed in this paper. {\it Lower panels:}
Window spectra of the data shown in the {\it upper panels}. The central peak is shown in detail. Amplitudes are given in arbitrary units.}
\label{time-d}
\end{figure}

\Section{The Data, Reductions and Analysis}
The data analyzed in this paper were obtained in the ASAS3-North station located at Haleakala (Maui, Hawaii Islands) using two 
wide-field instruments, equip\-ped with Nikkor 200-mm f/2.0 lenses and Apogee AP-10, 2048\,$\times$\,2048 CCD cameras, 
collecting data in two filters, $V$ and $I$. The data cover roughly 500 days between July 2006 and December 2007.
Distribution of observations in time for an arbitrarily chosen star in the Kepler FoV is shown in Fig.~\ref{time-d}. 
Typically, only one point per two or even four nights 
was obtained during the first year of observation; the frequency of observations in $I$ was increased to several points 
per night in the second half of 2007. This way of observing results in window spectra shown in Fig.~\ref{time-d}. 
They are characterized by strong daily aliasing, as expected.  On average, the $V$-filter data set consisted of $\approx$90 measurements, 
while about 110 data points were available in the $I$-filter. 

While searching for variability, we must also consider the problem of correct identification of variable
stars. In the wide-field surveys, the frames have low spatial resolution which for the ASAS frames amounts 
to about 15$^{\prime\prime}$/pixel. The photometry in these frames is extracted through five different apertures 
with diameters ranging from 2 to 6 pixels corresponding to 0.5$^\prime$ to 1.5$^\prime$. More details of image 
processing and reductions can be found in the previous ASAS papers (Pojma\'nski 1997, 2002). In quite dense areas
of the Kepler FoV, the low spatial resolution of ASAS frames poses obvious problems with contamination by nearby stars
in the extracted photometry.  In some cases, only follow-up observations with better resolution can help in proper 
identification of a variable star.  We will return to this problem in Section 4.

The magnitudes of stars measured in the ASAS frames range from 7 mag to 15 mag in $V$ and from 6 mag to 14 mag in $I$, thus very well
overlap with the range for Kepler targets. The observed field is located at intermediate Galactic latitudes 
($+$6{\deg} $< b <$ $+$23{\deg}).
Both magnitude range and the location of the field in the sky has a direct consequence as to the variability that can be detected.
For example, only a small number of early-type stars can be expected in the field.

The variability search was carried out using photometry in the intermediate aperture (diameter of 4 pixels) by means of 
Fourier amplitude periodogram calculated in the range between 0 and 30~d$^{-1}$. All light curves and periodograms were 
inspected visually. The variability type was assigned taking into account the period, amplitude and/or the shape of 
the light curve. In total, 947 stars were selected as variable out of about 250\,000 searched for variability. 

\Section{Description of the Catalog}
The catalog is available at:\\
\centerline{\it http://www.astro.uni.wroc.pl/ldb/asas/kepler.html}
and its mirror in the ASAS project web page:\\
\centerline{\it http://www.astrouw.edu.pl/asas/?page=kepler}

\begin{figure}[!h]
\includegraphics[width=12.5cm]{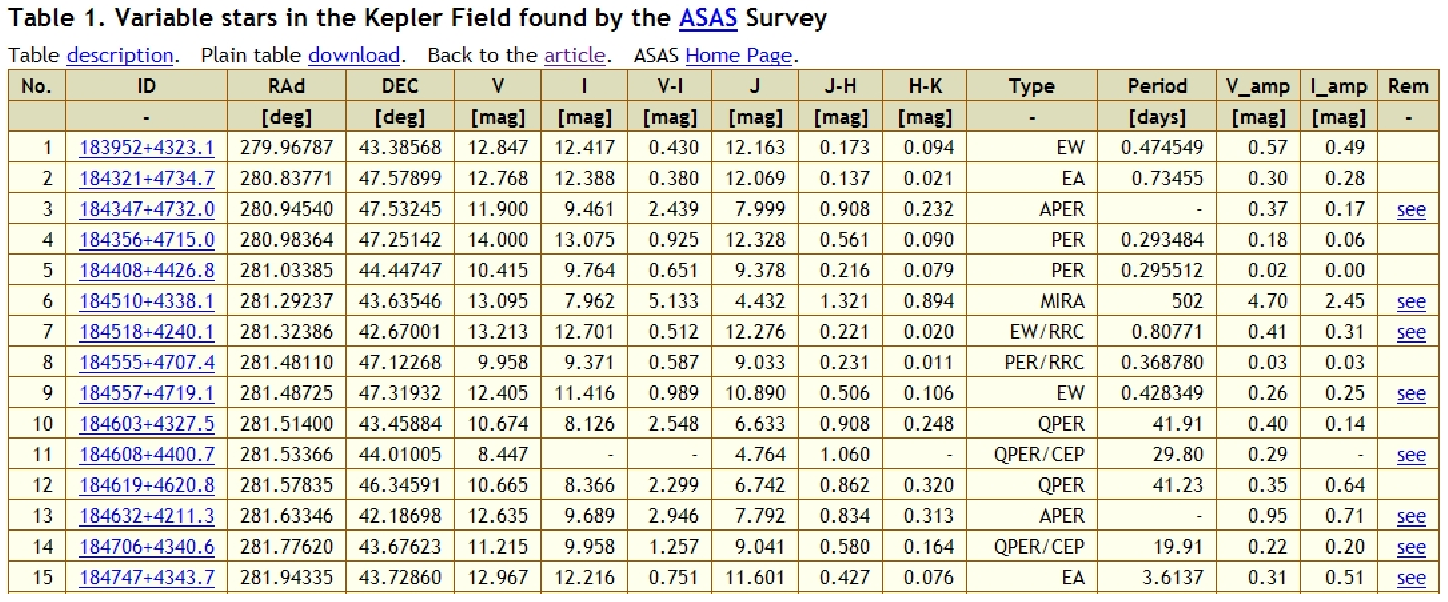}
\FigCap{The layout of the table containing catalog entries.}
\label{tab1}
\end{figure}

The main part of the catalog is the table that can be accessed and retrieved only through the web page; a detailed description of
its contents can also be found there. The layout of the table is shown in Fig.~\ref{tab1}. 
The table contains: equatorial coordinates, $VI$ magnitudes from the ASAS, $JHK_{\rm s}$ photometry from 2MASS catalog of
point sources (Cutri \etal 2003) if a star was identified with a 2MASS source, type of variability, 
variability period or dominating period (if a star showed periodic
or quasiperiodic behavior), variability ranges in $V$ and $I$ and remarks. As far as the type of variability is concerned, we decided to assign
only eleven variability flags, although the catalog may include more types of variability. First of all, the eclipsing binaries
are divided into three groups according to the shape of the light curve following the ``General Catalog of Variable Stars'' (GCVS) definitions: 
Algol-type (EA), $\beta$~Lyr-type (EB) and W\,UMa-type (EW). Out of different types of pulsating stars, we distinguish only those
which are easily recognizable by their period, shape of the light curve and/or amplitude, namely: RR Lyr stars of Bailey type 
ab and c (RRAB, RRC), high-amplitude $\delta$~Sct stars (HADS), Cepheids (CEP) and Miras (MIRA). The remaining stars are divided 
into three groups depending on how well periodicity is defined. Stars classified as PER are strictly periodic with sinusoidal light
curves. Those with QPER variability flag have a dominating periodicity, but show also amplitude and/or phase changes or 
periodic changes superimposed on the variability on a longer time scale. Stars of this type are commonly referred to as 
semi-regular or long-period variables. Finally, stars that do not show any well-pronounced periodicity or the period is longer
than the interval covered by the analyzed data, are designated as APER.

\begin{figure}[!hb]
\includegraphics[width=12.5cm]{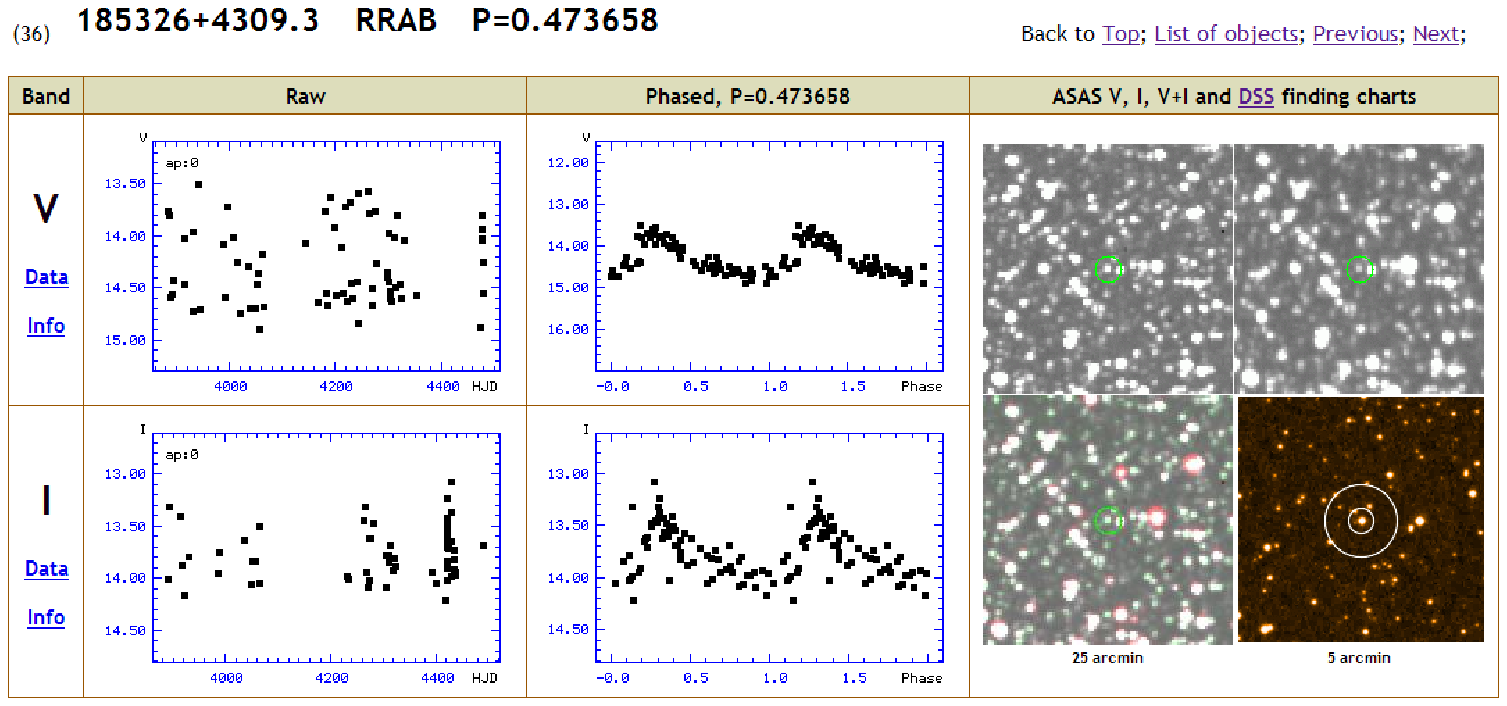}
\FigCap{A sample graphical interface to ASAS data, light curves and finding charts (see text for explanation).}
\label{maps}
\end{figure}

The link on the left-hand side of the table with ASAS ID opens a new page (Fig.~\ref{maps}) containing a header with sequential number, 
ASAS ID, type of variability and period (if available). On the right-hand side of the header the links to the first (Top), next and 
previous objects in the list as well as to the full table are available.

Below the header there are panels that show different plots in a form of a table with four columns, two rows and a header. 
The first row contains data in $V$, the second, in the $I$-band. The links to the original ASAS data (Data) and short information 
on the format (Info) are given in the first column. The second column contains light curves plotted as a function of heliocentric 
Julian Day. The data are shown for aperture that displayed the smallest scatter. The scatter depends on the object brightness: 
aperture ``0'' (2 pixels wide) is used for $V >$ 14 mag or $I >$ 12~mag and larger apertures for brighter stars, up to aperture ``4'' (6 pixels 
wide) for $V <$ 9 mag or $I <$ 7~mag. The aperture number is given in the plot.

For stars showing (quasi)periodic variability, the phase diagram is also shown in the third column if period was shorter than 150 days. 
For the remaining stars this column is left blank. Finally, the fourth column contains four finding charts, centered on the variable 
star position. The first three images, 25$^\prime$\,$\times$\,25$^\prime$, were extracted from the deep (stacked) ASAS $V$ and $I$ images. 
The third image is color-coded combination of $V$ (green) and $I$ (red) images, that can be useful to identify blends of stars 
having strongly different colors (e.g.~185650+4757.2 or 190015+3934.7). The fourth image is a 5$^\prime$\,$\times$\,5$^\prime$ 
cut from the Digitized Sky Survey.\footnote{The Digitized Sky Survey was produced at the Space Telescope Science
Institute under U.S.~Government grant NAG W-2166. The images of these surveys are based on photographic data obtained using the Oschin Schmidt
Telescope on Palomar Mountain and the UK Schmidt Telescope. The plates were processed into the present compressed digital form with the
permission of these institutions.} Two central circles correspond to the ASAS smallest (2 pixels, 30$^{\prime\prime}$ wide) and largest 
(6 pixels, 90$^{\prime\prime}$ wide) photometric apertures.

The table that follows graphics shown in Fig.~\ref{maps} (not shown in this figure) contains essentially the same entries as the main table of the catalog. 
The new entry is the 2MASS cross-identification (or a `BLEND' entry if the identification was not possible due to blending).
Also, the remarks, notes and cross-identifications are given explicitly there. Finally, the position of a star in the Kepler field 
(chip number and position in pixels) for four different orientations of the satellite is provided.
The cross-identifications in our catalog were taken from five other sources of
the information on variability in this area of the sky, namely: GCVS, ``New Catalog of Suspected Variable Stars'', 
including the Supplement (Kazarovets \etal 1998 -- star numbers from this catalog are preceded by ``NSV''), ROTSE1 catalog 
(Akerlof \etal 2000), Northern Sky Variability Survey (NSVS, Wo\'zniak \etal 2004a), red variables (Wo\'zniak 
\etal 2004b) and RR Lyr stars (Wils \etal 2006) and Hungarian-made Automated Telescope Network (HATnet) 
catalog (Hartman \etal 2004). The names from the latter catalog are preceded by ``HAT199''.

\Section{Variable Stars}
The quality of measurements combined with their number resulted in a detection threshold (in the Fourier periodogram)
ranging from 0.01~mag to 0.05~mag for stars brighter than $\approx$11~mag in $I$. It was very similar in the $V$-filter, for stars brighter 
than $\approx$12~mag. This large range in accuracy is mainly the result of different crowding; in general, in less crowded
fields the accuracy was better. For fainter stars the detection threshold grows rapidly reaching about 0.15~mag for the faintest stars ($\approx$15~mag
in $V$ and 14~mag in $I$). This means that our catalog is quite strongly biased towards variable stars that have large amplitudes. 
We present below a short summary of the contents of the catalog regarding different types of variable stars emphasizing particularly
different pulsating stars.  We must, however, stress that the classification we provide is not always unique because the period, 
amplitude and shape of the light curve were sometimes insufficient to do this unambiguously. For this reason some stars were given double 
or even triple variability flag, the most probable going first.
\MakeTable{cccc}{10cm}{Number of stars in the catalog with different variability flags assigned}
{\hline
\noalign{\vskip1pt}
Variability& Number of stars & Already & New \\
 flag (VF) & with a given VF & known & \\
\noalign{\vskip1pt}
\hline
\noalign{\vskip1pt}
CEP & 6  & 3 & 3\\
RRAB & 11 & 9 & 2\\
RRC & 9 & 3 & 6 \\
HADS & 4 & 2 & 2 \\
EA & 51 & 26 & 25\\
EB & 12 & 4 & 8 \\
EW & 123 & 35 & 88 \\
PER & 75 & 12 & 63 \\
QPER & 404 & 166 & 238 \\
APER & 194 & 93 & 101 \\
MIRA & 58 & 52 & 6 \\
\noalign{\vskip1pt}
\hline
\noalign{\vskip1pt}
Total & 947 & 405 & 542 \\
\noalign{\vskip1pt}
\hline
\label{vtab}
}

Table 1 summarizes the contents of our catalog listing the numbers of stars with a given variability flag. In case of multiple
classification, only the first one was considered. Of 947 stars we have in the catalog, 542 ($\approx$57\%) are new detections. The location of 
all variable stars in the Kepler FoV we present here is shown in Fig.~\ref{map}.

\begin{figure}[htb]
\includegraphics[width=11cm]{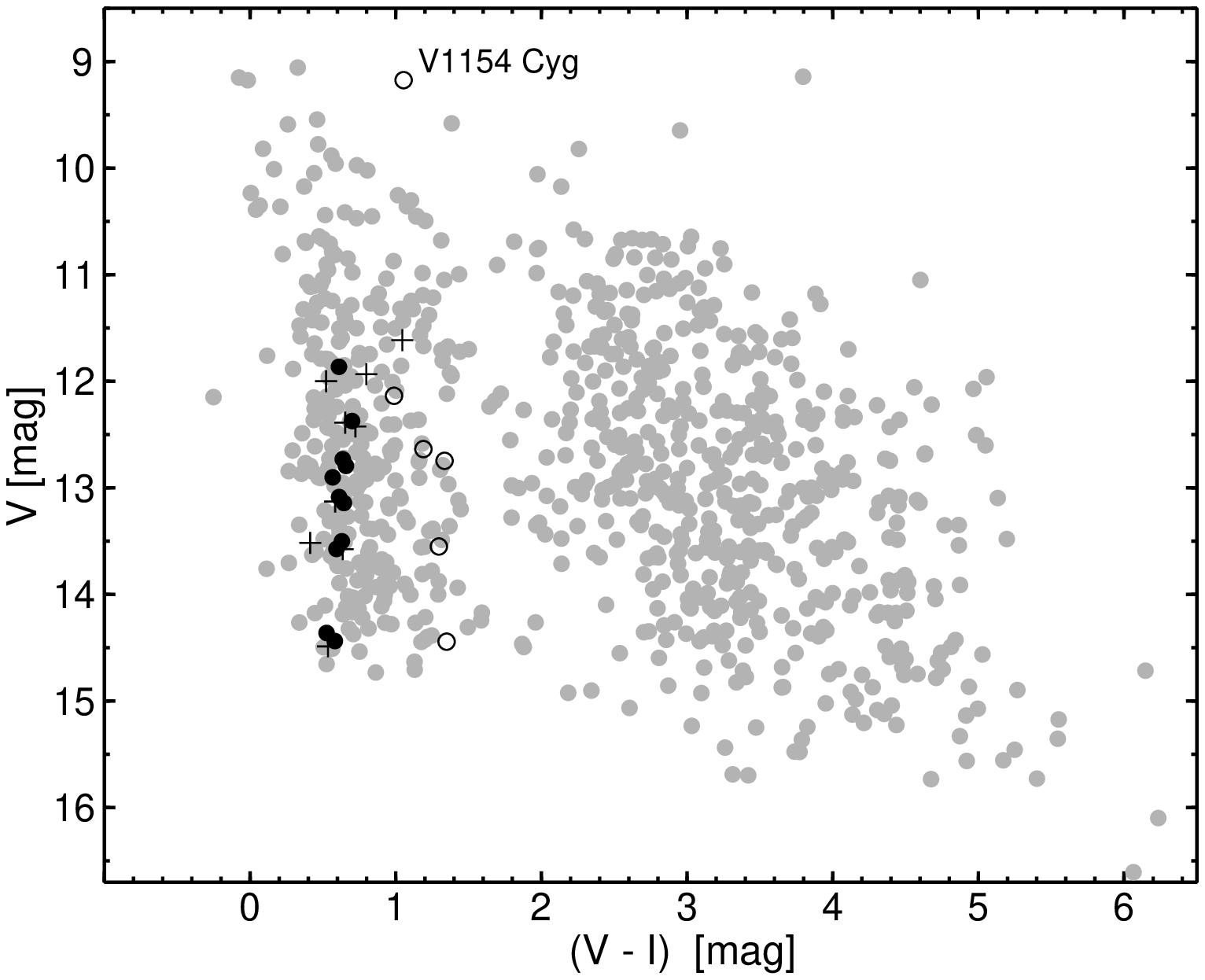}
\FigCap{The color-magnitude, $V$ vs.~($V-I$), diagram for variable stars from our catalog (gray symbols) with V1154\,Cyg (labeled), candidate 
Cepheids (open circles) and RR Lyrae stars of Bailey type ab (filled circles) and c (plus signs) plotted.}
\label{cep-rr}
\end{figure}

The color-magnitude diagrams for variable stars from our catalog are shown in Figs.~\ref{cep-rr}--\ref{lpv}. As can be seen in these
figures and as was pointed out above, the
$V$ magnitudes of these stars fall very well within the range of magnitudes that will be covered by the Kepler satellite. Two groups of stars can be
seen in these diagrams. The stars with $(V-I) <$ 1.6~mag are mainly main-sequence stars while the stars with redder colors are mainly red giants.

\subsection{Cepheids}
The only well-known Cepheid in the Kepler FoV is V1154~Cyg = BV\,394 = ASAS\,194815+4307.6 = \#721\footnote{The number preceded by 
a ``\#'' denotes sequential number of a star from our catalog.} ($V$ = 9.2~mag) discovered by Strohmeier (1962).
It has a period of 4.93~d. There are 18 stars in our catalog that have ``CEP'' variability flag assigned, but only six, including V1154\,Cyg, 
were given this flag as the first one. These six stars are shown in Fig.~\ref{cep-rr} as open circles. Not counting V1154\,Cyg, they should 
be regarded as candidates for Cepheids. The best of them seem to be ASAS\,192345+5116.2 (\#322, $P$ = 13.37~d) with $V$ magnitude range clearly larger
than $I$, as expected for Cepheids, and ASAS\,192152+4911.9 ($P$ = 3.4521~d). However, a new multi-color photometry and/or spectroscopy is
highly desirable to verify their variability type, as alternative explanations (e.g.~chromospherically active stars) are also possible for these
stars. The other stars, if verified as Cepheids, are much more distant than V1154\,Cyg (see Fig.~\ref{cep-rr}).

\subsection{RR Lyr Stars}
RR Lyr stars, especially of RRab type, are well recognizable by their non-sinusoidal light curves, large amplitudes and periods of 0.4--0.8 d. 
There are 11 RRab stars in our catalog, but only two, \#24 and \#320, are new findings. The remaining ones were discovered in previous searches,
mainly ROTSE (Akerlof \etal 2000) and NSVS (Wils \etal 2006). The RRc stars have smaller amplitudes and shorter periods than RRab ones. 
In consequence, faint stars (that is, with a less accurate photometry) of this type are not well
distinguishable from eclipsing binaries of W\,UMa type that have equally deep minima. We have 19 stars with ``RRC'' flag in the catalog, but only 
nine (six are newly discovered) have this flag given as the first one. The nine most certain RRc and eleven RRab stars are plotted in 
Fig.~\ref{cep-rr}. As can be seen, except for one RRc star (which might be misclassified EW star) they cover quite a narrow range of ($V-I$) 
color. Since extinction and reddening in the Kepler FoV are rather low, their distances reach $\approx$8 kpc. These are not the farthest RR\,Lyrae 
stars in the field as there are $\approx$30 known RRab stars in the Kepler FoV with $V >$~15 mag. We do not have them in our catalog because 
they were too faint to be detected in the ASAS data.

\subsection{Main-Sequence Pulsators}
Main-sequence pulsators ($\beta$~Cep, SPB, $\delta$~Sct and $\gamma$~Dor stars) are potentially very good targets for asteroseismology as 
they usually have many modes excited. Unfortunately, these modes have typically very small amplitudes, not exceeding several mmag, and therefore
hard to detect in our data.  The modes observed in the main sequence pulsators have sinusoidal light curves and therefore the
shape of the light curve does not help in the classification. There is one exception: evolved $\delta$~Sct stars that usually show only radial
modes having large photometric amplitudes. They are called high-amplitude $\delta$~Sct (HADS) stars. We have four HADS stars in our 
catalog, two are new findings. One of them, \#717 = ASAS\,194803+4146.9, is the most interesting one as it shows double-mode behavior with period
ratio equal to 0.763, typical for radial first overtone to fundamental period ratio (see, e.g., Pigulski \etal 2006).
\begin{figure}[htb]
\includegraphics[width=11cm]{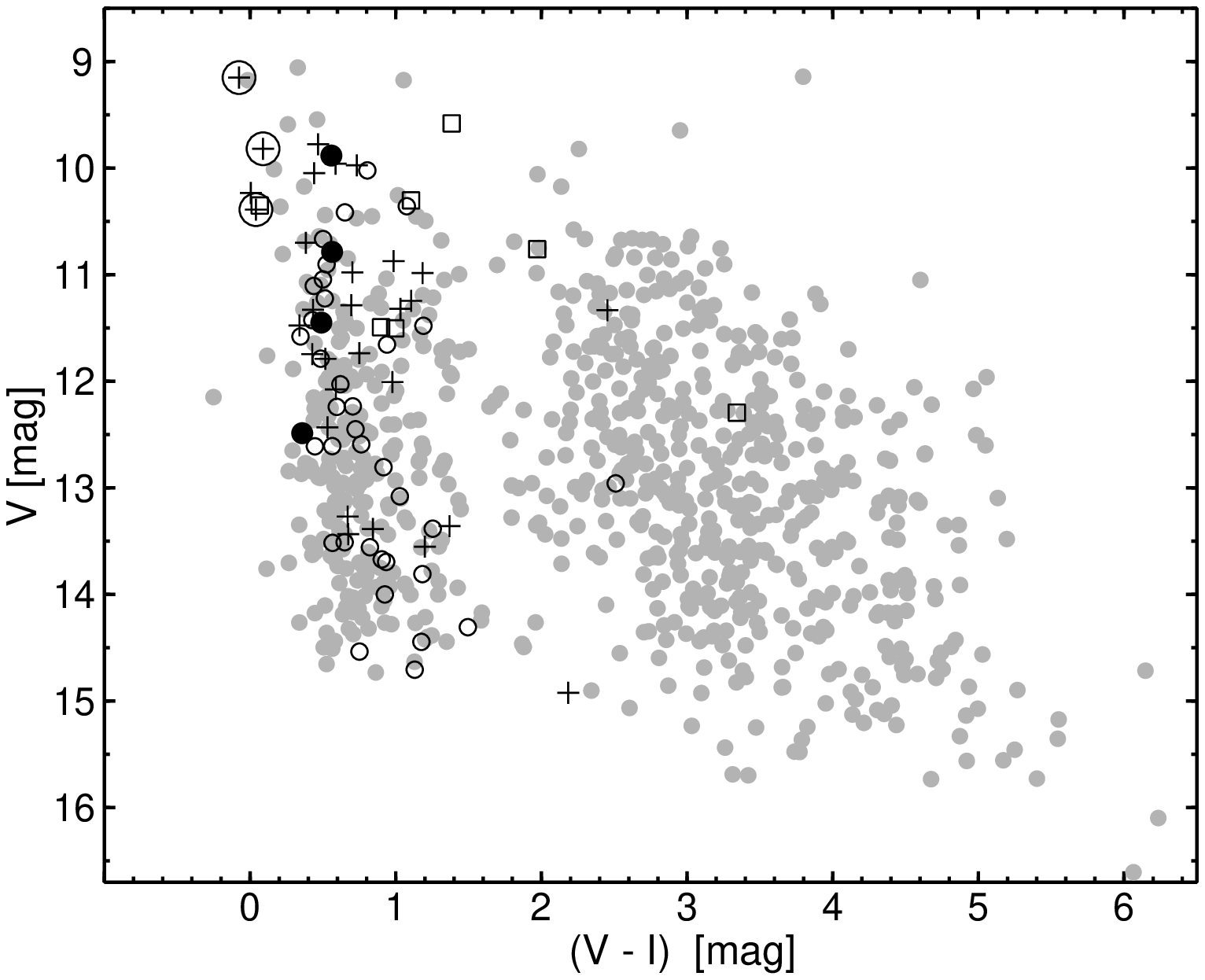}
\FigCap{The same as in Fig.~\ref{cep-rr}, but for the candidates for main-sequence pulsators. The following groups of
stars were distinguished: HADS stars (filled circles), stars classified as periodic variable (``PER'') with periods shorter than 0.3~d (open circles),
with periods between 0.3 and 5 days (plus signs), and periods longer than 5 days (open squares). Three likely SPB stars are shown as 
encircled plus signs.}
\label{msp}
\end{figure}

Not counting the HADS stars, the main sequence pulsators can be sought in our catalog among stars with a ``PER'' flag. We have divided these stars 
into three groups according to the period that we detected: (i) stars with periods shorter than 0.3~d (39 stars), (ii) stars with periods in the range 
between 0.3 and 5~d (28 stars), and (iii) stars with periods longer that 5~d (8 stars). The 0.3-day limit divides the main sequence pulsators into 
the short-period ($\beta$~Cep and $\delta$~Sct) and long-period (SPB, $\gamma$~Dor) stars. Theoretically, the former stars pulsate mainly 
with acoustic
($p$) modes, the latter, with gravity ($g$) ones. Stars included in group (iii), if their variability is at all due to pulsations, are rather not main 
sequence stars. Distinguishing $\beta$~Cep from $\delta$~Sct stars or SPB from $\gamma$~Dor is, however, rather impossible using only 
the ASAS photometry. Additional information like spectral type would be conclusive in this case.

There is, in fact, a little chance that any star in our catalog is a $\beta$~Cep-type pulsator. One reason has been already explained;
this is large-amplitude bias in our catalog. The other reason is that Kepler FoV is located at intermediate Galactic latitudes. On the other hand,
$\beta$~Cep-type stars are intrinsically bright, with absolute magnitudes $M_{\rm V} < -$2.5~mag.  If a $\beta$~Cep star is nearby, 
it will be saturated in the ASAS images. If not, it has to be located at a large distance from the Galactic plane. This, in turn, is
rather unlikely as these stars are massive and young and except for runaway stars are expected to be located near star-forming regions, i.e., 
in the Galactic disk. In consequence, $\beta$~Cephei stars in the Kepler FoV can be expected only in those parts of this field which are 
located closest to the Galactic plane where the absorption is high and distant stars are still quite close to the Galactic plane. 
The short-period variable stars in our catalog, shown in Fig.~\ref{msp} with open
circles, are therefore mainly $\delta$~Scuti stars which are intrinsically fainter than $\beta$~Cep stars. 

Next, it has to be noted that some short-period stars in our catalog classified as `PER', might be spurious detections. The reason for this 
is the distribution of 
data, already commented in Section 2. For the ASAS data, the consecutive data points are typically separated by 2--4 days. The corresponding Nyquist frequency
is therefore very low, much lower than the pulsation frequencies of $\delta$~Scuti stars.  Fortunately, the data points are not evenly distributed 
in time. In consequence, the Nyquist frequency loses its meaning and we are able to indicate the correct frequency of pulsation (the ambiguity 
due to daily aliases remains though) in the Fourier amplitude spectrum. However, some consequences of this distribution of data remain.
One of them is that spurious high frequencies may sometimes appear above of the adopted threshold in the Fourier periodograms of the ASAS data.
For this reason, we initiated a follow-up program carrying out multi-color photometry aimed at verifying the short-period variability of stars in the 
Kepler FoV. The result of this program will be published elsewhere.

There are very few stars among those shown in Fig.~\ref{msp} that have $UBV$ photometry or spectral types available which can be used to make the final
classification of the variability. A spectroscopic low-resolution follow up of these stars would be highly desirable as the four types of main sequence 
pulsators under consideration can be easily separated using the observed periods and spectral types. For several stars, however, the MK spectral types
were already obtained. In particular, two stars can be reliably classified as SPB stars. The first one is HD\,176562 = \#71 = ASAS\,185900+4115.9 
($P$ = 0.76366~d)
classified as B6\,Vs by Dworetsky \etal (1982). The color indices provided by the same authors ($B-V$ = $-$0.11, $U-B$ = $-$0.50) also indicate that
it is unreddened mid-B star. The other SPB star is BD\,$+$43{\deg}3223 = \#293 = ASAS\,192135$+$4403.0 ($P$ = 1.10146~d). This star is 
classified by MacRae (1952) as B8. A few other stars have spectral types in Harvard classification indicating that they are likely SPB stars.
The best example is HD\,226795 = \#896 = ASAS\,195725$+$4038.1 ($P$ = 1.7015~d) which was given spectral type B8 by Cannon (1925). The three likely
SPB stars are shown in Fig.~\ref{msp} with encircled symbols. As can be seen, they are among the brightest and bluest objects in our catalog.

\begin{figure}[htb]
\includegraphics[width=11cm]{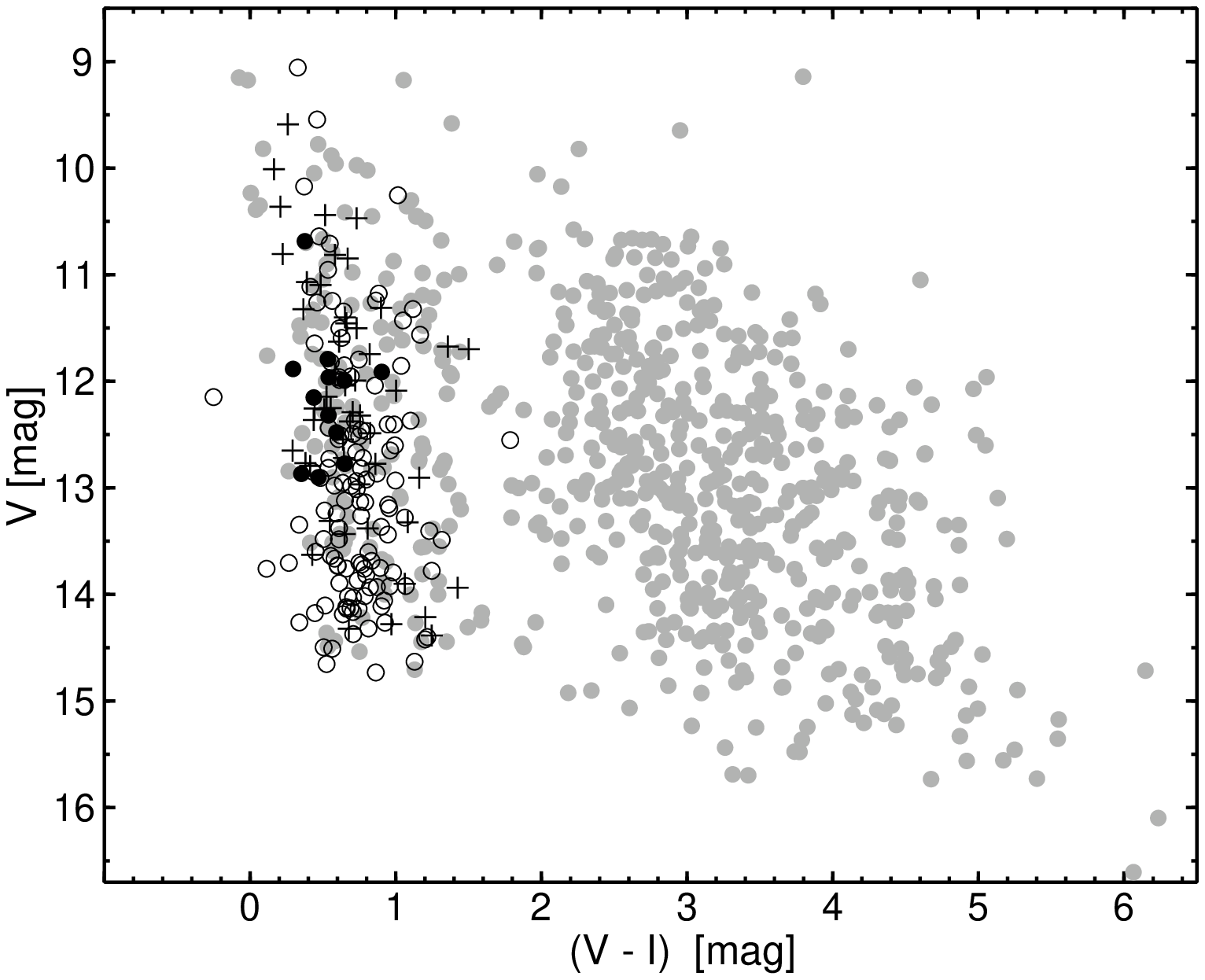}
\FigCap{The same as in Fig.~\ref{cep-rr}, but for eclipsing binaries classified as EA (plus signs), EB (filled circles), and EW (open circles).}
\label{ecl}
\end{figure}

\subsection{Eclipsing Binaries}
There are over 180 eclipsing binaries in our catalog, mostly classified as EW, i.e., presumably contact binaries of W UMa type. They are shown 
in Fig.~\ref{ecl}. As expected, they are mostly located in the left-hand part of this diagram because the components (at least the primaries) 
of these binaries are mostly main sequence stars. The three brightest eclipsing binaries in Fig.~\ref{ecl} are new variables: 
\#231 = ASAS\,191638$+$4805.8 ($V$ = 9.06), \#129 = ASAS\,190700$+$4632.3 ($V$ = 9.54) and \#289 = ASAS\,192112$+$4758.7 ($V$ = 9.59).

\subsection{Long-Period Variable Stars}
\begin{figure}[htb]
\includegraphics[width=11cm]{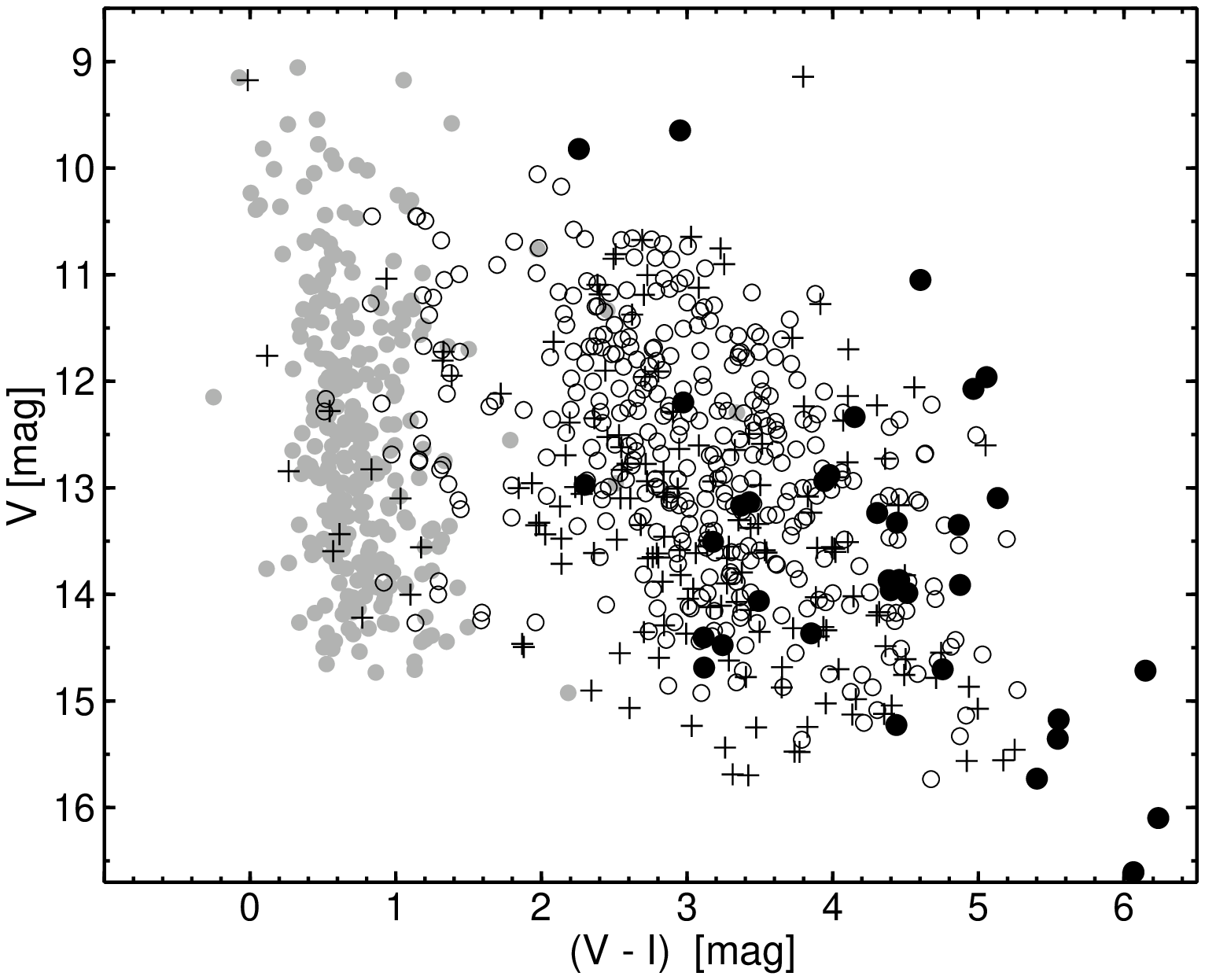}
\FigCap{The same as in Fig.~\ref{cep-rr}, but for long-period variables: Miras (filled circles), quasi-periodic (open circles) and aperiodic (plus
signs) stars.}
\label{lpv}
\end{figure}

The long-period variable stars, i.e.~stars which we classified as APER, QPER and MIRA form the most numerous group of stars in our catalog.
All but few stars with ($V-I$) $>$ 1.6~mag (Fig.~\ref{lpv}) belong to this group. The location of Miras in Fig.~\ref{lpv} is not well defined
because we use simply a mean in each filter which is not good for variable with such large amplitudes as Miras.

While Miras (58 in our catalog) are easily recognizable due to their large amplitudes, this is not the case for the remaining 
long-period variable stars.  These stars are predominantly red giants or supergiants.  They have recently been systematized 
by Soszy\'nski \etal (2007) using OGLE observations of Magellanic Clouds and diagrams where Wesenheit indices are plotted {\it vs.}~log(period). Owing to the small absorption towards the Magellanic Clouds and nearly the same distance of stars in each of them, 
red variables form well-defined sequences in these diagrams.  They can be in most cases explained by different types of pulsations 
and/or binarity. Unfortunately, similar diagrams for the the long-period variable stars in our catalog does not form such sequences
because they are smeared effectively by different distances of stars.

\section{Conclusions}
The catalog we make available in the present paper is a response for the need to have an homogeneous variability survey 
in the whole Kepler FoV that could be used in the selection of targets for the Kepler satellite, especially for the asteroseismic part of its program. 
Due to the limited number of observations gathered up-to-date in the ASAS3-North station, the catalog is quite strongly biased towards 
high-amplitude variable stars. The observations of the Kepler FoV are continued, however, and future analysis should allow to discover stars with
much smaller amplitudes than presented here. Next, some follow-up programs are needed to verify classification we made, especially for the
main-sequence pulsators (see Section 4.3). Both multi-color photometry and low-resolution spectroscopy would be desirable.  The CCD photometry
with telescope(s) having much better spatial resolution than the ASAS3 equipment would also allow to decide in many cases which star is variable.
The need for such a photometry comes from the fact that contamination by nearby stars might be in many cases a severe problem and affect the aperture 
photometry which is obtained from the ASAS frames. Finally, the areas located relatively close to the Galactic plane should be monitored because
probability of finding hot pulsators is the largest in this part of the Kepler FoV.  The first follow-up program of this kind has already been 
started. Results of these observations will be presented in a separate paper.

\Acknow{We are indebted to Mr.~Wayne Rosing who has kindly allocated space and provided 
technical support of the LCOGT staff for ASAS-North instruments inside The Faulkes Telescope North on Haleakala, 
Maui, Hawaii. We thank Dr.~David Koch who kindly provided his RA2fpix program used to calculate the
positions of stars in the Kepler CCD chips. This project was supported by the N2030731/1328 and N\,N203\,302635 grants from the MNiSzW.}

\end{document}